\documentclass[12pt]{article}
\usepackage{graphicx}

\begin{document}
\title{ANCOVA: A HETEROSCEDASTIC GLOBAL TEST WHEN THERE IS CURVATURE AND
 TWO COVARIATES}
\author{Rand R. Wilcox \\
Dept of Psychology \\
University of Southern California\\
}
\maketitle
\pagebreak
\begin{center} 
ABSTRACT
\end{center}



For two independent groups, let $M_j(\mathbf{X})$ be some conditional measure
 of location for the $j$th group associated with some random variable $Y$ given  $\mathbf {X}=(X_1, X_2)$.  
Let $\Omega=\{\mathbf{X}_1, \ldots, \mathbf{X}_K\}$ be a set of $K$ points to be determined.
An extant technique can be used to test $H_0$: 
 $M_1(\mathbf{X})=M_2(\mathbf{X})$ for each $\mathbf{X} \in \Omega$
without making any parametric assumption about $M_j(\mathbf{X})$. 
But there are two general reasons to suspect that the method can have relatively low power.
 The paper reports simulation results on an alternative approach that is designed  to test the global hypothesis
$H_0$: $M_1(\mathbf{X})=M_2(\mathbf{X})$ for all $\mathbf{X} \in \Omega$.
The main result is that the new method offers a distinct power advantage. 
 Using data
from the Well Elderly 2 study, it is illustrated that the alternative method can make a 
practical difference in terms of detecting a difference between two groups.

Keywords:  
ANCOVA, trimmed mean, smoothers, Well Elderly 2 study

\section{Introduction}

For two independent groups,  consider the  situation  where for the $j$th group ($j=1$, 2)
$Y_j$ is  some outcome variable of interest  and $\mathbf{X}=(X_{1}, X_{2})$ is a vector of two covariates. 
Let  $M_j(\mathbf{X})$ be some conditional 
robust measure of location associated with $Y$ given $\mathbf{X}$.
A basic and well known
  goal is determining whether the groups differ in terms of $M_1(\mathbf{X})$ and $M_2(\mathbf{X})$. 
 The classic  ANCOVA (analysis of covariance) method assumes that 
\begin{equation}
Y_j = \beta_{0j} + \beta_1X_{1j} + \beta_2X_{2j} + \epsilon, \label{usual}
\end{equation}
where $\beta_{0j}$,  $\beta_1$ and $\beta_2$ are unknown parameters estimated via least squares regression
 and $\epsilon$ is a random variable having a normal distribution with mean zero and unknown variance $\sigma^2$.
So the regression planes are assumed to be parallel and the goal is to compare the 
intercepts.
It is well known, however, that there are serious concerns with this approach. 
First, there is a vast literature establishing that methods based on means, and more broadly 
least squares regression, are not robust (e.g.,  Staudte and Sheather, 1990; 
 Marrona et al., 2006; 
Heritier et al., 2007; Hampel et al., 1986;
Huber and Ronchetti, 2009;  Wilcox, 2012). 
 A  practical consequence is that power can be relatively low
even under a small departure from normality. 
 Moreover, even a single outlier can yield  a poor fit to the bulk 
 of the points when using least squares regression. 
Another concern is that two types of homoscedasticity are assumed. The first
is that for each group, the variance of the error term does not depend on the value of the covariate. 
If this assumption is violated the wrong standard error is being used. 
The second is that the variance of the error term is the same for both groups.
Violating these assumptions can result in poor control over the Type I error probability. 
Yet another fundamental concern with (\ref{usual}) is that the true regression surfaces are assumed to be planes. 
Certainly, in some situations, this is a reasonable approximation.
 When there is curvature, using some obvious parametric regression model might suffice. 
 (For example, include a 
quadratic term.) But it is known that this approach can be inadequate, 
which has led  to a substantial collection of nonparametric regression methods, 
often called smoothers, for dealing with curvature in a more
flexible manner
(e.g., H\"{a}rdle, 1990;  Efromovich, 1999;  Eubank , 1999; Fox, 2001; Gy\"{o}rfi, et al., 2002).  
Yet another concern is the assumption that the regression surfaces are parallel. 
One could test the assumption that the slope parameters are 
equal, but it is unclear when such a test has enough power to detection situations 
where this assumption is violated to the point that it makes
a practical difference. 


Here, the model given by (\ref{usual})  is replaced with the less restrictive model
 \begin{equation}
Y_j = M_j( \mathbf{X}) + \epsilon_j, \label{model}
\end{equation}
where  $M_j(\mathbf{X})$ is some unknown  function that reflects some conditional 
robust measure of location associated with $Y$ given 
 $\mathbf{X}$. The random
variable  $\epsilon_j$ has some unknown distribution with variance $\sigma^2_j$. 
So unlike the classic approach where it is assumed that 
\[ M_j(\mathbf{X}) =  \beta_{0j} + \beta_{1j}X_1 + \beta_{2j}X_2, \]
no parametric model for $M_j(\mathbf{X})$ 
 is specified and  $\sigma^2_1=\sigma^2_2$ is not assumed.
In particular it is not assumed that the regression surfaces are parallel. 
The goal here is  to test  the global hypothesis
\begin{equation}
H_0: M_1(\mathbf{X})=M_2(\mathbf{X}), \, \forall \, \mathbf{X} \in \{\mathbf{X}_1, \ldots, \mathbf{X}_K\}, \label{null_glob}
\end{equation}
where $\mathbf{X}_1, \ldots, \mathbf{X}_K$ are $K$ vectors chosen empirically  in a manner  to be determined. 
 
  For the case of a single covariate,
Wilcox (2012, section 11.11.1) describes a method that tests $H_0$: $M_1(X_k)=M_2(X_k)$
for each $k$, $k=1, \ldots, K$.
Roughly, for each $X_k$, identify values of the covariate that are close to $X_k$ and then 
compare the groups based on the corresponding $Y$ values using a method based on a robust measure of location. 
For this special case, it is a relatively simple matter to choose values for the covariate in a manner that is likely to find any differences that might
exist.

When dealing with two covariates, Wilcox (2012) suggests a simple  extension where the values  of the covariate are chosen based on
how deeply they are nested within the cloud of covariate values. 
(This is method M1 in section 2.1 of this paper.)
The $K$ points are chosen to include the point in the first group having
the deepest half space depth plus the points on the .5 depth contour. (More precise details are given in section 2.)
This typically results in a relatively small number of covariate values where the corresponding $Y$ values are compared based on 
 a robust measure of location. Again $K$ tests are performed and the probability of one or more
Type I errors can be controlled using some improvement on the Bonferroni method (e.g., Rom, 1990; Hochberg, 1988).
 But it is not al all clear when this relatively simple approach will choose
covariate values that are likely to detect true differences between the groups. 
A way of dealing with this issue is to select a larger collection of covariate values, but if  the
 familywise error rate (the probability of one or more Type I errors) is controlled, power can be relatively poor due to the large number of
 hypotheses that are tested. 
 Switching to a method that controls the false discovery rate  when  dealing with dependent test statistics (e.g., 
 Benjamini \&Yekutieli, 2001) would suffer from the same concern.
 So the focus here is on testing (\ref{null_glob}) using a specified proportion of the deepest covariate values within 
 the cloud of covariate values that are available. 
 
 The paper is organized as follows. Section 2 reviews the method in Wilcox (2012) followed by a description of an alternative method aimed at
 testing (\ref{null_glob}). Two variations of the alternative method
 are compared via simulations in Section 3 in terms of both power and their ability to control the Type I error probability.  The 
 power of both variations is compared to the power of the method in Wilcox (2012). Section 4 uses data from the Well Elderly 2 study to illustrate
 that the new method can make a practical difference.
 
 \section{Description of the Methods}
 
The methods compared here are based in part on a method derived by 
 Yuen (1974) for comparing the
population trimmed means of two
independent groups. To describe it, momentarily ignore the covariates and consider the goal of testing
\begin{equation}
H_0: \mu_{t1}=\mu_{t2},
\end{equation}
the hypothesis that two independent groups have equal population trimmed means. 
For the $j$th group ($j=1$, 2), let $n_j$ denote the sample size
and let $Y_{(1)j} \le \ldots \le Y_{(n_j)j}$ denote the $Y_{ij}$ values written
in ascending order.  For some  $0 \le \gamma <.5$, the $\gamma$-trimmed mean for the $j$th 
 group is
\[\bar{Y}_j = \frac{1}{n-2g_j}\sum_{i=g_j+1}^{n-g_j} Y_{(i)j},\]
where $g_j=[\gamma n_j]$ is the greatest integer less than or equal to $\gamma n_j$. 
Here the focus is on $\gamma=.2$, a 20\% trimmed mean. Under normality, this choice has good efficiency relative to the sample mean
(Rosenberger \& Gakso, 1983). Moreover, the sample 20\% trimmed mean  enjoys certain theoretical advantages. First, it has a reasonably 
high breakdown point, which refers to the proportion of values that must be altered to destroy it.  Asymptotic results and simulations 
indicate that it reduces substantially  concerns about the impact of skewed distributions on the probability of a Type I error. 
This is not to suggest that 20\% trimming  is always the optimal choice: clearly this is not the case. 
The only suggestion is that it is a reasonable choice among the many robust estimators that might be used. 

Winsorizing the $Y_{ij}$ values  refers to  setting
\begin{equation}
 W_{ij} = \left\{ \begin{array}{ll}
          Y_{(g+1)j}, & \mbox{if $Y_{ij} \le Y_{(g_j+1)j}$} \\
          Y_{ij}, & \mbox{if $Y_{(g_j+1),j} < Y_{ij} < Y_{(n-g_j)j}$} \\
          Y_{(n_j-g_j)j}, & \mbox{if $Y_{ij} \ge Y_{(n-g)j}$.}
            \end{array}
    \right.
\end{equation}
The  Winsorized sample mean  corresponding to group $j$ is
\[\bar{W}_j = \frac{1}{n_j} \sum W_{ij},\]
and the Winsorized variance is 
\[s^2_{wj} = \frac{1}{n_j-1} \sum (W_{ij}-\bar{W}_j)^2. \]
 Let
$h_j=n_j-2g_j$. That is, $h_j$ is the number of observations left
in the $j$th group after trimming. Let
\begin{equation}
d_j = \frac{(n_j-1)s^2_{wj}}{h_j(h_j-1)}.
\end{equation}
Yuen's test statistic is
\begin{equation}
T_y = \frac{\bar{X}_{t1} - \bar{X}_{t2}}{\sqrt{d_1 + d_2}}. \label{yuen}
\end{equation}
The null distribution is taken to be a Student's t distribution with 
 degrees of freedom 
\[\hat{\nu}_y = \frac{(d_1+d_2)^2}{\frac{d_1^2}{h_1-1}+\frac{d_2^2}{h_2-1}}.\]

\subsection{Method M1}

Method M1 is described in Wilcox (2012, section 11.11.3). A complete description of the  many computational details is not provided
here, but an outline of the method is provided with the goal of explaining how it differs from method M2 in the next section.

Let $\mathbf{X}_{ij}$ ($i=1, \ldots n_j$; $j=1$, 2) denote the $n_j$ covariate points corresponding to the $j$th group.
Momentarily consider a single covariate point, $\mathbf{X}$.  Method M1 estimates $M_j(\mathbf{X})$ using the 
$Y_{ij}$ such that the corresponding $\mathbf{X}_{ij}$ values are close to ${\mathbf X}$. More precisely, 
$\mathbf{X}_{ij}$ ($i=1, \ldots n_j$; $j=1$, 2)for the $j$th group,
compute  a robust covariance matrix based on $\mathbf{X}_{ij}$ ($i=1, \ldots, n_j$). 
There are many ways of computing a robust covariance matrix with no single estimator dominating. Here a skipped covariance
matrix is used, which is computed as follows. 
For fixed $j$, outliers among the  $\mathbf{X}_{ij}$ values are identified using a projection-type multivariate outlier detection technique
(e.g., Wilcox, 2012, section 6.4.9). These outliers are removed and the usual covariance matrix is computed using the remaining data.

 Next, compute robust Mahalanobis distances for each covariate point based on the
robust covariance matrix  just described, with $\mathbf{X}$ taken to be the center of the data. The point $\mathbf{X}_{ij}$
 is said to be close to  $\mathbf{X}$
if its robust  Mahalanobis distance is small, say less than or equal to $f$, which is called the span. Generally $f=.8$ performs reasonably well
when the goal is to approximate the regression surface. Of course exceptions are encountered, but henceforth $f=.8$ is assumed. 
Let $P_j(\mathbf{x})$ be the subset of \{1, 2, \ldots, $n_j$\} that indexes  the $\mathbf{X}_{ij}$ values
 such that the Mahalanobis distance associated with  $\mathbf{X}_{ij}$  is less than or equal to $f$.
Let $N_j(\mathbf{X})$ be the cardinality of the set $P_j(\mathbf{X})$ and let  $M_j(\mathbf{X})$ denote the  20\% trimmed mean  based on the 
$Y_{ij}$ values for which $i \in P_j(\mathbf{X})$.  
Then for the single point  $\mathbf{X}$, (\ref{null_glob}) can be  tested by applying Yuen's method with the $Y_{ij}$ values
for which  $i \in P_j(X)$ provided  both $N_1(X)$ and $N_2(X)$ are not too small. Following Wilcox (2012),
 this is taken to mean that Yuen's method can be applied if  simultaneously $N_1(X) \ge 12$ and $N_2(X) \ge 12$, in which
case the two groups are said to be comparable at $\mathbf{X}$.

Now consider the issue of choosing covariate values where the regression surfaces will be compared.  For fixed $j$,
compute how deeply each $\mathbf{X}_{ij}$  is nested within the cloud of points $\mathbf{X}_{ij}$
 ($i=1, \ldots, n_j$).
This is done with a projection type method that is similar to an approach discussed by Donoho and Gasko (1992).
Computational details are described in section 2.2. Consider the deepest point as well as those on the polygon
containing the central half of the data.  (Liu et al.,  1999, call  this polygon the .5 depth contour.) Method M1 applies Yuen's method at  each
of these points provided the regression surfaces are comparable at these points as previously defined.
The probability of one or more   Type I errors is controlled using the method in Hochberg  (1988). 

\subsection{Method M2}

There are several positive features of method M1 but some negative features as well. First, 
Yuen's method for comparing trimmed means has been studied extensively and appears to perform relatively well in terms of both Type I errors
and power. The method for  choosing the covariate values  seems reasonable in the sense that it uses
points that are nested deeply within the cloud of covariate points, which  reflect situations where the regression surfaces are comparable. 
Roughly, deeply nested points correspond to situations where the regression surfaces can be estimated in a relatively accurate manner. 
If a point $\mathbf{X}$
is not
deeply nested in the cloud of covariate values, finding a sufficiently large number of other points that are close to $\mathbf{X}$ might
be impossible. 

But a concern about M1 is that perhaps true differences might be missed because of the relatively small number of covariate values
that are used. A way of dealing with this possibility is to use all of the covariate points that are deeply nested in the cloud of all
covariate points and then test the global hypothesis given by (\ref{null_glob}). This is the strategy behind method M2.

Method M2 begins by computing the projection depth (e.g., Wilcox, 2012, section 6.2.5) 
for each $\mathbf{X}_{i1}$ (the $i$th covariate vector in group 1) in the same manner as method M1.
To describe the computational details, momentarily focus 
 on a single $n \times p$ matrix of data, ${\bf Z}$. Let $\hat{\tau}$ be some robust measure of location based on 
${\bf Z}$. For simplicity, the marginal medians (based on the usual sample median) are used. 
 Let
\[{\bf U}_i = {\bf Z}_i - \hat{\tau}\]
($i=1, \ldots, n$), 
\[
C_i=  {\bf U}_i {\bf U}_i^{\prime}.
\]
For any $j$ ($j=1, \ldots, n$),  let
\[V_{ij} = \sum_{k=1}^J U_{ik} U_{jk},\]
\begin{equation}
T_{ij} = \frac{V_{ij}}{C_i} (U_{i1}, \ldots, U_{ip}) \label{bigt}
\end{equation}
and
\[D_{i \ell} =  \| T_{ij}\|,\]
where $\| T_{ij}\|$ is the Euclidean norm associated with the
vector $T_{ij}$ ($i=1,\ldots n$; $j=1, \ldots, n$).
Let 
\[d_{ij} = \frac{D_{i \ell}}{q_{i2}-q_{i1}},\]
where $q_{i2}$ and $q_{i1}$ are estimates of the upper and lower quartiles, respectively,  based on 
$D_{i1}, \ldots, D_{in}$. (Here, $q_{i2}$ and $q_{i1}$ are based on the so-called ideal fourths; see Friqqe et al., 1989.)
The projection distance of ${\mathbf Z}_j$,  the $j$th row of ${\mathbf Z}$, relative to the cloud of points represented by  ${\mathbf Z}$,
 is the maximum value of $d_{ij}$, say $p_d({\mathbf Z}_j)$, the maximum being taken over $i=1, \ldots, n$ (cf. Donoho \& Gasko, 1992).
 Following Liu et al.  (1999), the depth of ${\mathbf Z}_j$ is taken to be 
 \[P_D({\mathbf Z}_j)=\frac{1}{1+p_d({\mathbf Z}_j)}.\]
 
Let  the set  $\{ \mathbf{X}_1, \ldots, \mathbf{X}_K\}$  indicate the deepest half  of the points in the first group. 
Points where the regression surfaces are not comparable (i.e.,  $N_1(\mathbf{X}) < 12$ or $N_2(\mathbf{X}) < 12$) 
are discarded. Because $K$ can be relatively large, controlling 
FWE via Hochberg's method seems likely to have relatively low power, which is verified in the simulations in section 4. 

The reason for choosing the deepest half, rather than some larger proportion, is based on preliminary simulations. Using the deepest half, typically
the regression surfaces are comparable at all $K$ points when  the sample sizes for both groups are greater than or equal to 50.
For a larger proportion of points, this is often not the case. There are, of course, many other variations. Some other measure of the depth
might be used or one could use  all of the covariate points where the regression surfaces are comparable. The goal here is to find at least
one variation that controls the Type I error probability reasonably well and simultaneously offers a power advantage over method M1.

Method M2 begins  in the same manner as method M1: test $H_0$: $M_1(\mathbf{X})=M_2(\mathbf{X})$ for each
 $\mathbf{X} \in \{\mathbf{X}_1, \ldots, \mathbf{X}_K\}$. Label the
resulting p-values $p_1, \ldots,  p_K$. The idea is to test  (\ref{null_glob}) using some function of these $K$ p-values. 
Perhaps the best-known method for  testing some global hypothesis based on p-values  is a technique derived by Fisher (1932).
But Zaykin et al. (2002) note that  the ordinary Fisher product test loses power in cases where there are
a few large p-values. They suggest using instead a truncated product method (TPM), which is based on the test statistic
\[W = \prod_{k=1}^K  p_k^{I(p_i \le \tau)}\]
where $I$ is the indicator function. Setting $\tau=1$ yields Fisher's method, but Zaykin et al. suggest using $\tau=.05$. 
Zaykin et al. derive the null distribution of $W$ when all $K$ tests are independent. But  the $K$ tests performed here are not 
independent simply because $P_j(\mathbf{X}_k) \cap P_j(\mathbf{X}_{\ell})$, $k \ne \ell$, is not empty. If this dependence among the tests is ignored when 
computing a critical value for $W$, control over the Type I error probability is poor. For the dependent case, Zaykin et al. suggest using
a bootstrap method, but this results in relatively high execution  time for the situation at hand making this approach difficult to study via simulations. 
Consequently, an alternative approach was used: Proceed as done by Gosset in his derivation of Student's t and assume normality 
with the goal of determining  the $\alpha$ quantile of $W$, say $w$, in which case (\ref{null_glob}) is rejected
at the $\alpha$ level if $W \le w$. 
Here, the critical value $w$ was  determined  via simulations
 using (2) with $M_j(\mathbf{X}) \equiv 0$ and $\epsilon_j$ having a
standard normal distribution.   More precisely, for each $j$,
$(Y_{ij}, \mathbf{X}_{ij})$ ($i=1, \ldots n_j$; $j=1$, 2)   were generated 
from a trivariate normal distribution where all correlations are zero. Then $W$ was computed and this
process is repeated say $B$ times yielding $W_1, \ldots, W_B$. Put these $B$ values
in ascending order yielding $W_{(1)}  \le \ldots \le W_{(B)}$. Then $w$ was estimated to be
$W_{(k)}$, where $k$ is $\alpha B$ rounded to the nearest integer. Here, $B=4000$ was used.

One of many alternative methods is to use instead the test statistic 
\[\bar{Q} = \frac{1}{K} \sum _{k=1}^K  p_k.\]
Unexpectedly, this alternative test statistic performed relatively well, in terms of power, under a shift in location model, as illustrated in section 4. 
Now reject (\ref{null_glob})  if $\bar{Q} \le q_{\alpha}$, the  $\alpha$ quantile of $\bar{Q}$, which again
is determined via simulations in the same manner as the critical value $w$.


\section{Simulation Results}

As is evident, 
 a basic issue is the impact on the Type I error probability when dealing with non-normal distributions as well as situations where there is
an association with the covariate variables. 
Simulations were used to address this issue with $n_1=n_2=50$.  Smaller sample sizes, such as $n_1=n_2=30$,
routinely result in situations where no covariate values can be found where comparisons can be made.
That is, $N_1(\mathbf{X}) < 12$ or $N_2(\mathbf{X}) < 12$ for all $\mathbf{X} \in  \{\mathbf{X}_1, \ldots, \mathbf{X}_K\}$.
 
Estimated Type I error probabilities, $\hat{\alpha}$, were based on 4000 replications.
Four types of distributions were used:
normal, symmetric and heavy-tailed, asymmetric and light-tailed,
and asymmetric and heavy-tailed.
More precisely, values for the error term, $\epsilon_j$ in (\ref{model})  were generated from  one of four g-and-h distributions
(Hoaglin, 1985) that contain the standard  normal distribution as a special case. 
If $Z$ has a standard normal distribution, then by definition
\[V = \left\{ \begin{array}{ll}
 \frac{{\rm exp}(gZ)-1}{g} {\rm exp}(hZ^2/2), & \mbox{if $g>0$}\\
  Z{\rm exp}(hZ^2/2), & \mbox{if $g=0$}
   \end{array} \right. \]
has a g-and-h distribution where $g$ and $h$ are parameters that
determine the first four moments. 
The four distributions used here were the standard normal ($g=h=0$), a
symmetric heavy-tailed distribution ($h=0.2$, $g=0.0$), an asymmetric
distribution with
relatively light tails ($h=0.0$, $g=0.2$), and an asymmetric distribution with
heavy tails ($g=h=0.2$).
Table 1 shows the skewness ($\kappa_1$) and kurtosis
($\kappa_2$)
for each distribution. Additional properties of the g-and-h distribution
are summarized by Hoaglin (1985).
The $\mathbf{X}_{ij}$ values were generated from a bivariate normal distribution  with correlation equal to zero. 

\begin{table}
\caption{Some properties of the g-and-h distribution.}
\centering
\begin{tabular}{ccrr} \hline
g & h &  $\kappa_1$ & $\kappa_2$\\ \hline
0.0 & 0.0  & 0.00 & 3.0\\
0.0 & 0.2 & 0.00 & 21.46\\
0.2 & 0.0  & 0.61 & 3.68\\
0.2 & 0.2 & 2.81 & 155.98\\ \hline
\end{tabular}
\end{table}

Three types of associations were considered. The first two deal with situations where $Y_{ij}=\beta X_{ij}+ \epsilon$. The two choices for the slope
were $\beta=0$ and 1.  The third  type was
$Y_{ij}= X_{ij}^2+ \epsilon$.  These three situations are labeled S1, S2 and S3, respectively.  
 Additional simulations were run where the  correlation between the two covariates is .5.
But this had almost no impact on the results, so for brevity they are not reported. 

Estimated Type I error probabilities are reported in Table 2.
Although the seriousness of a Type I error depends on the situation, 
Bradley (1978) suggests that as a general guide, when testing at the .05 level, the actual level should
be between .025 and .075. Based on this criterion, both TPM and  the method based on $\bar{Q}$ provide adequate control
over the Type I error probability. A possible appeal of TPM is that
 when testing at the .05 level, the actual level was estimated to be less than or equal to .050 among all of the situations considered.
 As for the method based on $\bar{Q}$, the estimate exceeds .05 in some situations, particularly when dealing with 
 heavy-tailed distributions ($h=.2$), the largest estimate being .069.


\begin{table}
\center
\caption{Estimated  Type I error probabilities when testing at the $\alpha=.05$ level, $n_1=n_2=50$}
\begin{tabular}{ccc cc}
$g$ & $h$ &   S &  $\bar{Q}$ &  TPM \\ \hline
0.0 & 0.0   &    1 &   .050 &   .050 \\
0.0 & 0.0   &    2 &   .036  & .042 \\
0.0 & 0.0   &    3 &   .048  & .049 \\
0.0 & 0.2   &    1 &   .061 &  .046 \\
0.0 & 0.2   &    2 &   .050  &  .043 \\
0.0 & 0.2   &    3 &   .064   & .048 \\ 
0.2 & 0.0   &    1 &   .055 &  .046 \\
0.2 & 0.0   &    2 &   .042  &  .038 \\
0.2 & 0.0   &    3 &   .052  & .046 \\
0.2 & 0.2   &    1 &   .064  & .047 \\
0.2 & 0.2   &    2 &   .053 & .042 \\
0.2 & 0.2   &    3 &   .069 & .048\\ 
\hline
\end{tabular}
\end{table}

Table 3 shows the estimated power when for the first group, (\ref{model}) is replaced by 
$Y_1 = M_1( \mathbf{X}) + \epsilon_1+.5$. As is  evident, method M2 based on 
$\bar{Q}$ has the highest power among all of the situations considered 
and method M1 has the lowest power. For some situations, the higher power using $\bar{Q}$, rather than TPM,
is presumably due in part to a lower Type I error probability associated with TPM. Note, however, that even in situations where both methods
have similar Type I error probabilities, $\bar{Q}$ has a higher estimated power. 

\begin{table}
\center
\caption{Estimated  power, $n_1=n_2=50$}
\begin{tabular}{ccc ccc}
$g$ & $h$ &   S &  $\bar{Q}$ &  TPM  & M1 \\ \hline
0.0 & 0.0   &    1 &  .409  & .345 &    .318\\   
0.0 & 0.0   &    2 &  .332  & .301&     .262\\ 
0.0 & 0.0   &    3 &   .341 & .307 &  .270\\ 
0.0 & 0.2   &    1 &   .387 & .290  &.283 \\                    
0.0 & 0.2   &    2 &   .299 & .245   & .229\\ 
0.0 & 0.2   &    3 &   .315 & .255  & .239\\ 
0.2 & 0.0   &    1 &   .410 & .327 & .324\\  
0.2 & 0.0   &    2 &   .327 & .303  & .276\\ 
0.2 & 0.0   &    3 &   .342 & .287 & .270\\ 
0.2 & 0.2   &    1 &  .388 & .286 & .284 \\  
0.2 & 0.2   &    2 &  .299 & .243 & .231\\ 
0.2 & 0.2   &    3 &   .318 & .247 & .232\\                  
\hline
\end{tabular}
\end{table}

Some additional simulations were run where for the first group, $Y_j = M_j( \mathbf{X}) + \epsilon_j+.5I_{X_1>0}$. The idea was that 
the two versions of method M2 are a function of the pattern of the individual p-values and that perhaps a situation where a difference
between the two regression surfaces exists only for a subset of the covariate values might result in TPM having higher power than
$\bar{Q}$. But again, $\bar{Q}$ had higher power than TPM. However, results in  section 4 indicate that in practice, 
$\bar{Q}$  does not dominate TPM in terms of power. 
















\section{Illustrations}

Data from the Well Elderly 2 study (Clark et al., 2011; Jackson et al., 2009) are used to illustrate that the choice of method
can make a practical difference. 
 A general goal in the Well Elderly 2 study was to assess the efficacy of an intervention strategy  aimed
at improving the physical and emotional health of older adults.   
A portion of the study
 was aimed at understanding the impact of intervention on a measure of meaningful activities 
 which was measured with the Meaningful Activity Participation Assessment (MAPA) instrument  (Eakman et al., 2010).
 Two covariates are used here. The first is a measure of
depressive symptoms based on the
  Center for Epidemiologic 
Studies Depressive Scale (CESD).  The CESD (Radloff, 1977) is sensitive to change in depressive 
status over time and has been successfully used to assess ethnically diverse older people (Lewinsohn et al., 1988; Foley et al., 2002). Higher scores indicate a higher level
of depressive symptoms.

The other covariate was the cortisol awakening response (CAR).
 Saliva samples were taken at four times over the course of a single day:  on rising, 30-60 minutes after rising, but before taking anything by mouth, before lunch, and before dinner. 
 Then samples were assayed for cortisol.
 Extant studies (e.g.,
 Clow et al., 2004; 
 Chida \& Steptoe, 2009) indicate that measures of
  stress are associated with the
 cortisol awakening response (CAR), which is defined as the change in cortisol concentration
that occurs during the first hour after waking from sleep. (CAR is taken to be the cortisol level upon awakening  minus the level of cortisol after the participants were awake for about an hour.) 
 The sample size for the control group was 187 and the sample size for the  group that received intervention was 228.
 Based on method M1, no significant differences were found with the familywise error rate set at .05. In contrast, method M2 based on
 $\bar{Q}$ rejects (the p-value is .008) and  the TPM version of method M2 rejects as well (the p-value is .021).
 
 Method M2 indicates that there is a difference between the two groups, but there is the issue of where and by how much. 
 A seemingly natural conclusion is that the groups differ at the point corresponding to the smallest p-value. 
 Here, the minimum p-value occurs for CAR equal to $-.218$ and CESD equal to  4.00, which correspond to a relatively high increase
 in cortisol after awakening coupled with a low CESD measure of depressive symptoms. Among the 74 covariate points that were used,
 39\% of the p-values are less than or equal to .05. 
More information can be gleaned from a plot of the p-values as well as the estimated difference between the regression surfaces where
comparisons were made.  Figure 1 shows a plot of the p-values for the situation at hand, which suggests that the strongest evidence for
a significant difference occurs when CESD is low regardless of what CAR might be.  

Figure 2
shows the estimated difference between the predicted MAPA scores. 
With one exception, all estimated differences are positive indicating that predicted MAPA scores are higher among the group receiving 
intervention. 
The highest estimated differences  occur for two subgroups of participants. The first consists of those  with a relatively high increase
 in cortisol after awakening coupled with a low CESD measure of depressive symptoms; the corresponding  p-values are relatively low.
The second subgroup consists of those  participants who have both a relatively high CAR and a relatively high CESD;
 these points have relatively low p-values as well. 
Among participants who have relatively high depressive symptoms and relatively low (negative) CAR values, the
difference between predicted MAPA scores is small. And as indicated in Figure 1, among these particular participants, highly non-significant
results were obtained.

 \begin{figure} 
\resizebox{\textwidth}{!}
{\includegraphics*[angle=0]{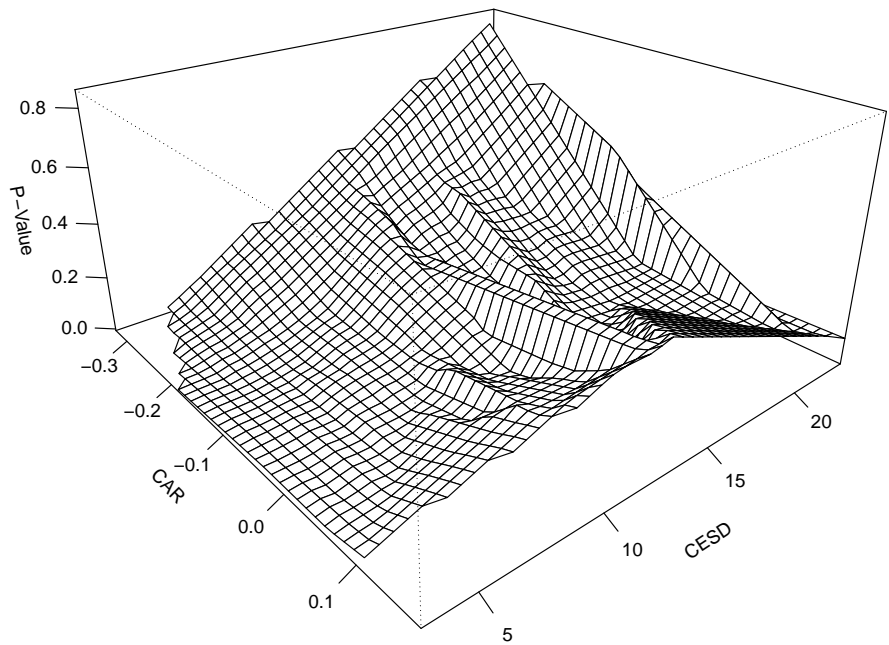}} 
\caption{The p-values associated with the covariate points where the regression surfaces were compared. The plot indicates that the
strongest evidence for a significant difference occurs when CESD is low.}
\end{figure}

 \begin{figure}
\resizebox{\textwidth}{!}
{\includegraphics*[angle=0]{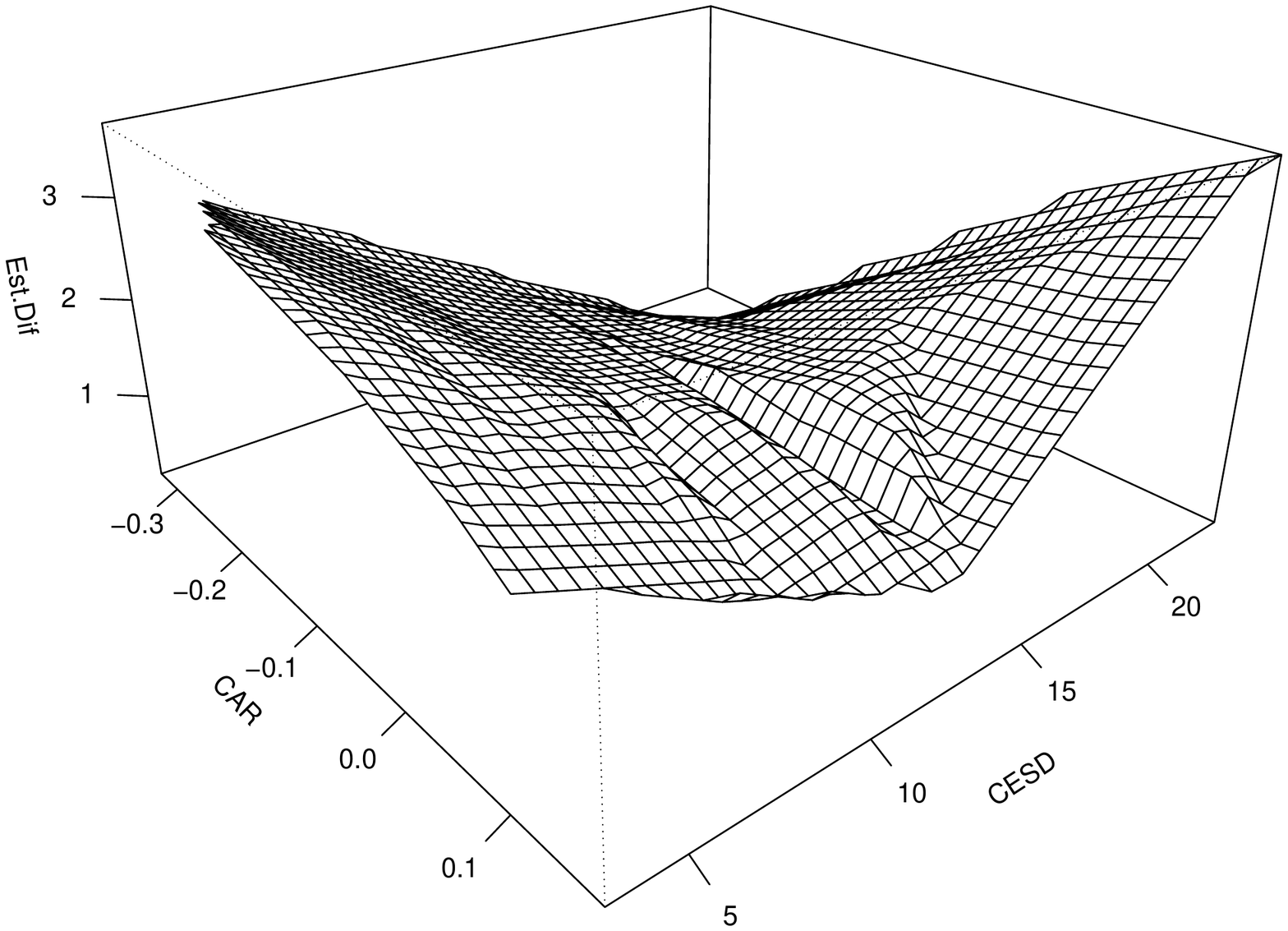}} 
\caption{The z-axis indicates the estimated difference between the predicted MAPA scores. 
(Positive values indicate higher predicted MAPA scores for participants in the intervention group.)
Estimated differences are relatively high  for 
two subgroups:   when the CAR is negative (cortisol increases shortly after awakening) and CESD is relatively low,  and when  both CESD and the CAR are relatively   high.}
\end{figure}


 Another dependent variable in the Well Elderly study
 was the RAND 36-item Health Survey (SF-36), a measure of self-perceived physical health and mental well-being 
(Hays, 1993; McHorney et al., 1993). Higher scores reflect greater 
perceived health and well-being. Here, the control group and the experimental
group are compared based on subset of the SF-36 items that reflect perceived physical health, again using CESD and the CAR as 
covariates. The deepest half of the data consisted of 74 covariate points.  Despite performing 74 tests, method M1 resulted in
a significant result for 6 of the 74 tests that were performed, again with FWE set equal to .05. The TPM version of Method M2 also rejects,
but $\bar{Q}$ does not reject; its p-value is .126.

 \section{Concluding Remarks}

In summary, all indications are that both versions of method M2 control the Type I error probability reasonably well. 
An apparent advantage of the TPM version of method M2 is that it avoids Type I error probabilities greater than the nominal level. But
 simulations indicate
that choice of method can make a practical difference in terms of power, with $\bar{Q}$ seeming to have an advantage. However,
the illustrations based on the Well Elderly 2 study suggest that $\bar{Q}$ does not dominate in terms of power. 

In principle, method M2 is readily extended to more than two covariates. 
But in practice this might require a relatively large sample size
due to the curse of dimensionality: neighborhoods with a fixed number of 
points become less local as the dimensions increase
(Bellman, 1961). 

There are many reasonable variations of method M2 and perhaps 
variations other than those studied here often provide a practical advantage.
For example, when using TPM, some other choice for $\tau$ might be 
more optimal in practice. In addition, there are many alternative test statistics that
might be used that are function of the individual p-values (e.g., Cousins, 2008).
As is evident, resolving this issue is non-trivial.

Finally, the R  function  ancov2COV, which is stored on the author's web page,
 performs both versions of method M2.



\begin{center}
REFERENCES
\end{center}

Bellman, R. E. (1961). {\em Adaptive Control Processes}.  Princeton, NJ: 
 Princeton  University Press.

Benjamini Y. \& Yekutieli D. (2001). The control of the false discovery rate in multiple testing under dependency. {\em Annals of Statistics, 29}, 1165--1188.

Bradley, J. V. (1978) Robustness? {\em British Journal of Mathematical and}  {\em Statistical Psychology, 31}, 144--152.



Chida, Y. \& Steptoe, A. (2009). Cortisol awakening response and psychosocial factors: A  systematic review and meta-analysis.
 {\em Biological Psychology, 80}, 265--278.

Clark, F., Jackson, J., Carlson, M., Chou, C.-P.,  Cherry, B. J.,  Jordan-Marsh M.,  Knight, B. G.,  Mandel, D.,  Blanchard, J., 
 Granger, D. A., Wilcox, R. R.,  Lai, M. Y.,  White, B.,  Hay, J.,  Lam, C., Marterella, A. \&  Azen, S. P. (2011). 
 Effectiveness of a lifestyle intervention in promoting
the well-being of independently living older people:
results of the Well Elderly 2 Randomise Controlled Trial.  {\em Journal of Epidemiology and Community Health, 66},
 782--790. doi:10.1136/jech.2009.099754


Clow, A., Thorn, L., Evans, P. \&  Hucklebridge, F. (2004). The awakening cortisol response: Methodological issues and significance.
{\em Stress, 7}, 29--37.

Cousins, R. D. (2008). Annotated bibliography of some papers on combining significances or  p-values. arXiv:0705.2209v2.



Donoho, D. L. \& Gasko, M. (1992). Breakdown properties of the location estimates  based on halfspace depth 
and projected outlyingness. {\em Annals of Statistics, 20}, 1803--1827.


Eakman, A. M., Carlson, M. E. \& Clark, F. A. (2010). The meaningful activity participation assessment: a measure
 of engagement in personally valued activities {\em International  Journal of Aging Human  Development, 70}, 299--317.

Efromovich, S. (1999). {\em Nonparametric Curve Estimation: Methods, Theory and 
 Applications}. New York: Springer-Verlag.


Eubank, R. L. (1999). {\em Nonparametric Regression and Spline Smoothing}. New York: Marcel Dekker.

Foley K., Reed P., Mutran E., et al. (2002). Measurement adequacy of the CESD  among a sample of older African Americans. {\em Psychiatric Research, 109}, 61--69.

Fox, J. (2001). {\em Multiple and Generalized Nonparametric Regression}.  Thousands Oaks, CA: Sage

Frigge, M., Hoaglin, D. C. \& Iglewicz, B. (1989). Some implementations of the  boxplot.  {\em American Statistician, 43}, 50--54


Gy\"{o}rfi, L., Kohler, M., Krzyzk, A. \&  Walk, H.  (2002). {\em A Distribution-Free Theory of Nonparametric Regression}. New York: Springer Verlag.

Hampel, F. R., Ronchetti, E. M., Rousseeuw, P. J. \& Stahel, W. A. (1986).
 {\em Robust Statistics}. New York: Wiley.

H\"{a}rdle, W. (1990). Applied Nonparametric Regression. Econometric
 Society Monographs No. 19, Cambridge, UK: Cambridge University Press.


Hays, R. D., Sherbourne, C .D. \& Mazel, R. M. (1993). The Rand 36-item health survey 1.0. {\em Health Economics, 2}, 217--227.

Heritier, S., Cantoni, E, Copt, S. \& Victoria-Feser, M.-P. (2009). {\em Robust Methods in Biostatistics}. New York: Wiley.



Hoaglin, D. C. (1985). Summarizing shape numerically: The g-and-h distribution. In D. Hoaglin, F. Mosteller \& J. Tukey (Eds.) {\em Exploring Data Tables Trends and Shapes}. New York: Wiley, pp. 461--515.

Hochberg, Y. (1988). A sharper Bonferroni procedure  for multiple tests of significance. {\em Biometrika, 75}, 800--802.

Huber, P. J. \& Ronchetti, E. (2009). {\em Robust Statistics}, 2nd Ed. New York: Wiley.

Jackson, J., Mandel, D., Blanchard, J., Carlson, M., Cherry, B., Azen, S., Chou, C.-P.,  Jordan-Marsh, M., Forman, T., White, B., Granger, D., Knight, B. \& Clark, F. (2009). Confronting challenges in intervention research with ethnically diverse older adults: the USC Well Elderly II trial. {\em Clinical Trials, 6}  90--101.



Lewinsohn, P.M., Hoberman, H. M., Rosenbaum M. (1988). A prospective study of risk factors  for unipolar depression. {\em Journal of  Abnormal Psychology, 97}, 251--64.

 
 Liu, R. Y., Parelius, J. M. \& Singh, K. (1999). Multivariate analysis by data depth: Descriptive statistics, graphics and inference. 
 {\em Annals of Statistics, 27}, 783--858.


 Maronna, R. A., Martin, D. R. \& Yohai, V. J. (2006). {\em Robust Statistics:}
 {\em Theory and Methods}. New York: Wiley.
 
 McHorney, C. A., Ware, J. E. \& Raozek, A. E. (1993). The MOS 36-item Short-Form Health Survey (SF-36): II. Psychometric and clinical tests of validity in measuring physical
 and mental health constructs. {\em Medical Care, 31}, 247--263.



Radloff, L. (1977). The CESD scale: a self report depression scale for research in the general population. {\em Applied Psychological Measurement 1}, 385--401.

Rom, D. M. (1990). A sequentially rejective test procedure based on a modified  Bonferroni inequality. {\em Biometrika, 77}, 663--666.

Rosenberger, J. L. \& Gasko, M. (1983). Comparing location estimators: Trimmed
 means, medians, and trimean. In D. Hoaglin, F. Mosteller and J. Tukey (Eds.) {\em Understanding Robust and exploratory data analysis.} (pp. 297--336). New York: Wiley.



Staudte, R. G. \& Sheather, S. J. (1990).
{\em Robust Estimation and Testing}.  New York: Wiley.


 Wilcox, R. R. (1997). ANCOVA based on comparing a robust measure of
 location at empirically determined design points. {\em British}
 {\em Journal of Mathematical and Statistical Psychology},
      {\em 50},  93--103.



Wilcox, R. R.  (2012).  {\em Introduction to Robust Estimation and Hypothesis Testing}, 3rd Edition. San Diego, CA: Academic Press.

Yuen, K. K. (1974). The two sample trimmed t for unequal population variances.
 {\em Biometrika, 61}, 165--170.

Zaykin, D. V., Zhivotovsky, L. A., Westfall, P. H., \& Weir, B. S. (2002). Truncated product method
for combining p-values. {\em Genetic Epidemiology 22}, 170--185.

\end{document}